\documentclass[aip,jcp,reprint,showkeys,superscriptaddress,a4paper,amsmath]{revtex4-1}
\usepackage{graphicx}
\usepackage{epsfig}
\usepackage{amsmath}
\usepackage{graphics}
\usepackage{latexsym}
\usepackage{amssymb}
\usepackage{inputenc}
\usepackage{color}
\usepackage{siunitx}
\usepackage[colorlinks=true,linkcolor=magenta,citecolor=cyan]{hyperref}

\newcommand{\la}{\left\langle}
\newcommand{\ra}{\right\rangle}

\begin{document}

\title{Structure and dynamics of amphiphilic patchy cubes in a nanoslit under shear}

\author{Takahiro Ikeda}
\affiliation{Faculty of Mechanical Engineering, Kyoto Institute of Technology, Goshokaido-cho, Matsugasaki, Sakyo-ku, Kyoto 606-8585, Japan}

\author{Yusei Kobayashi}
\email{kobayashi@kit.ac.jp}
\affiliation{Faculty of Mechanical Engineering, Kyoto Institute of Technology, Goshokaido-cho, Matsugasaki, Sakyo-ku, Kyoto 606-8585, Japan}

\author{Masashi Yamakawa}
\affiliation{Faculty of Mechanical Engineering, Kyoto Institute of Technology, Goshokaido-cho, Matsugasaki, Sakyo-ku, Kyoto 606-8585, Japan}

\begin{abstract}
Patchy nanocubes are intriguing materials with simple shapes and space-filling and multidirectional bonding properties. Previous studies have revealed various mesoscopic structures such as colloidal crystals in the solid regime and rod-like or fractal-like aggregates in the liquid regime of the phase diagram. Recent studies have also shown that mesoscopic structural properties, such as average cluster size $\la M \ra$ and orientational order, in amphiphilic nanocube suspensions are associated with macroscopic viscosity changes, mainly owing to differences in cluster shape among patch arrangements. Although many studies have been conducted on the self-assembled structures of nanocubes in bulk, little is known about their self-assembly in nanoscale spaces or structural changes under shear. In this study, we investigated mixtures of one- and two-patch amphiphilic nanocubes confined in two flat parallel plates at rest and under shear using molecular dynamics simulations coupled with multiparticle collision dynamics. We considered two different patch arrangements for the two-patch particles and two different slit widths $H$ to determine the degree of confinement in constant volume fractions in the liquid regime of the phase diagram. We revealed two unique cluster morphologies that have not been previously observed under bulk conditions. At rest, the size of the rod-like aggregates increased with decreasing $H$, whereas that of the fractal-like aggregates remained constant. Under weak shear with strong confinement, the rod-like aggregates maintained a larger $\la M \ra$ than the fractal-like aggregates, which were more rigid and maintained a larger $\la M \ra$ than the rod-like aggregates under bulk conditions. 
\end{abstract}
\maketitle

\section{Introduction}
Self-assembly is a ubiquitous phenomenon that occurs in colloidal suspensions. Because colloidal self-assemblies have unique physical properties, they have attracted interest in a wide range of research fields, such as drug delivery, optical sensors, and enhanced oil recovery. In addition, the transport properties of colloidal suspensions are affected not only by the properties of the constituent atoms, but also by their self-assembled structures. Therefore, the accurate control and prediction of self-assemblies are essential for the development of functional materials.

A promising approach to self-assembly manipulation is the modification of particle surfaces by introducing anisotropic surface interactions from patterned coatings (so-called ``patches'').\cite{zhang:nl:2004,wang:nat:2012,russo:rpp:2022} Directional bonding {\it via} surface patches provides diverse self-assembled, highly ordered structures, including chains,\cite{wang:nat:2012} micelles,\cite{srinivas:nl:2008,avvisati:jcp:2015} and network structures.\cite{walther:jacs:2009,russo:jcp:2009} In addition to surface anisotropy, particle-shape anisotropy plays a critical role in the formation of rich self-assembled structures.\cite{glotzer:nm:2007,vananders:an:2014,dwyer:sm:2023} The combination of surface and particle-shape anisotropies offers a wider range of possibilities for controlling self-assemblies and their resulting transport properties.

Among the various combinations of particle shape and surface anisotropy examined to date, patchy nanocubes have emerged as particularly intriguing materials owing to their simple geometry, space-filling properties, and multidirectional interaction anisotropy. For example, isotropic DNA-grafted nanocubes form a superlattice of colloidal crystals that exhibit a unique phase transition,\cite{knorowski:jacs:2014,lee:ns:2022,chen:sm:2023,zhang:jcp:2023} depending on the length and flexibility of the DNA strands. Recent studies of anisotropically patterned nanocubes in suspension based on molecular dynamics (MD) and Monte Carlo (MC) simulations\cite{kobayashi:la:2022} have revealed various finite-sized aggregates ranging from rods to fractal objects. A more recent study\cite{yokoyama:sm:2023} in which kinetic MC calculations were used to explore the behaviors of self-assembled clusters under shear found the shear-induced growth and breakup of these clusters. We also investigated the structural formation and rheological properties of amphiphilic cubes in bulk solutions at rest and under shear.\cite{ikeda:msde:2024}

Confinement in nanoscale channels is a key strategy for achieving novel self-assembled morphologies. Owing to the effect of spatial constraints and solid--liquid interfaces, confinement to nanoscale dimensions is known to induce unique structures and properties that differ from those in the bulk.\cite{lepri:prl:1997,gao:sr:2018,luo:pre:2021,jiang:jacs:2021,ishii:sa:2021,imamura:jcp:2024} Previous studies have extensively explored the structural dynamics of self-assembly in bulk systems; however, the impact of wall interactions has not received much attention. Some studies have demonstrated the distinctive self-assembled structures and morphologies of amphiphilic nanoparticles confined in narrow slit-like or tubular geometries. Fern{\'a}ndez {\it et al.}\cite{fernandez:pre:2014} found zigzag-like chain structures in Janus disks under quasi-one-dimensional confinement. Kobayashi and Arai\cite{kobayashi:sm:2015} found a variety of highly ordered structures in Janus nanoparticle suspensions confined in nanotubes, depending on the density and wall--colloid interactions. The authors further investigated the structural changes and rheological behaviors of these suspensions under nanotube flow by observing the change in shear-thinning rate associated with self-assembly rearrangement. Nikoubashman\cite{nikoubashman:sm:2017} studied the self-assembly of spherical Janus particles in microfluidic channels, and showed that the spatial cluster size distribution $M(z)$ changed from uniform to nonuniform with increasing flow rate. Baran {\it et al.}\cite{baran:jpcc:2020} observed different two-dimensional crystallizations of Janus spheres in two parallel solid plates of different widths. In a more recent study,\cite{baran:ao:2023} the authors reported a variety of layered structures parallel to the walls, depending on the density, slit width $H$, and particle--wall interactions. These studies show that complex self-assembled structures not observed in the bulk state occur even for common Janus spherical particles.

Because most previous simulation efforts have focused on Janus spheres in a confined geometry, knowledge about nonspherical particles such as amphiphilic nanocubes remains lacking. Unlike Janus spheres, amphiphilic nanocubes do not form micellar structures, which may be the origin of the peculiar structural morphologies not observed under bulk conditions, mainly owing to their intrinsic cluster shapes and wall--colloid interactions. Moreover, confined liquids are ubiquitous in experimental setups and natural phenomena; thus, further studies are necessary to provide valuable insights into the structural morphologies and resulting transport properties of nanocubes in realistic environments.

In this study, we investigated the structures and dynamics of amphiphilic nanocubes confined in a nanoslit geometry by incorporating repulsive interactions with walls at rest and under shear. To consider hydrodynamic interactions, we employed a combination of multiparticle collision dynamics (MPCD) and conventional MD.\cite{malevanets:jcp:1999} We investigated two mixtures of one- and two-patch cubes with different patch arrangements in a constant-volume fraction in the liquid regime of the phase diagram. We found that the cluster size of the rod-like aggregates depended on their degree of confinement $C$, whereas that of the fractal-like aggregates remained almost constant. 

\section{Model and methods}
\begin{figure*}
    \includegraphics[width=14.5cm]{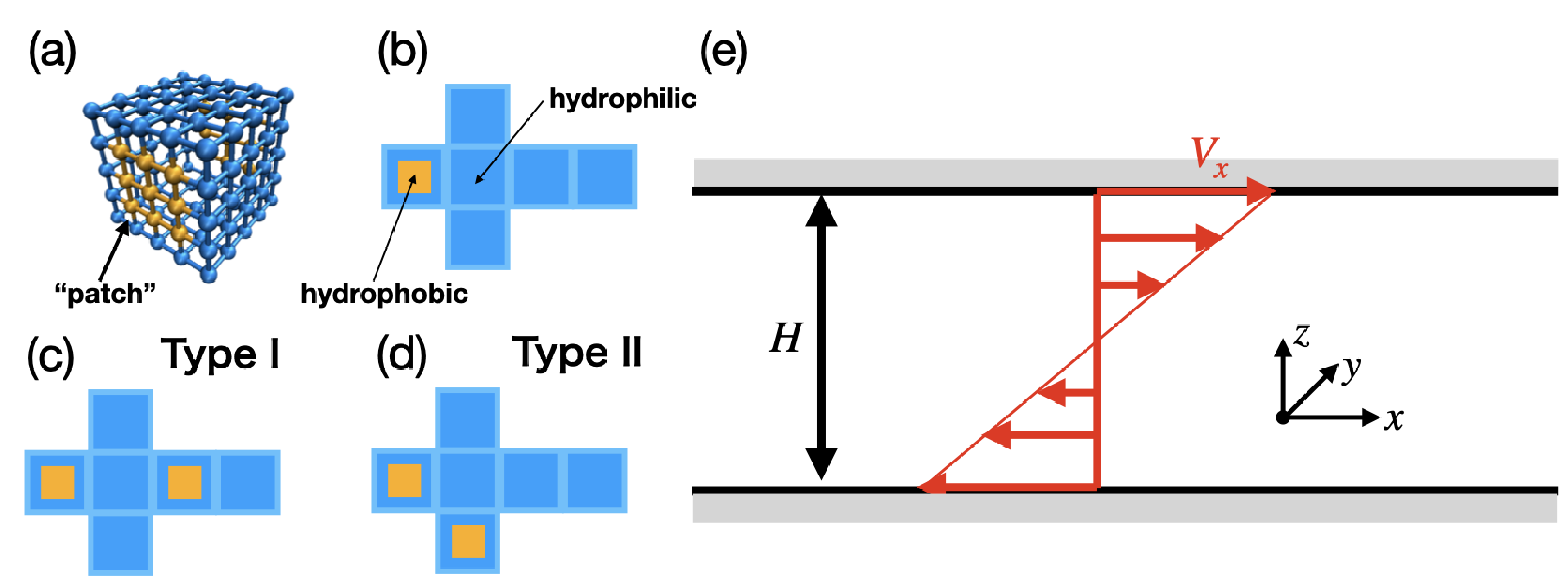}
    \centering
    \caption{(a) Discrete particle model of a cube with diameter $d=4\,a_{\rm v}$. The $N_{\rm v}$ vertex particles and bonds between nearest neighbors are shown (diametric bonds are omitted for clarity). (b--d) Unfolded representations of (b) a one-patch cube and (c, d) two-patch cubes. Hydrophilic and hydrophobic patches are colored blue and yellow, respectively. Snapshots were rendered using Visual Molecule Dynamics (version 1.9.4).\cite{vmd} (e) Schematic representation of a slit with width $H$. A uniform shear velocity profile is schematically depicted by red vectors. The maximum velocity in the flow direction is $\pm V_x$ at $\pm z=H/2$.}
    \label{fig:cubemodel}
\end{figure*}

We employ MD simulations coupled with the MPCD\cite{malevanets:jcp:1999,gompper:adv:2009,howard:coce:2019} technique to incorporate hydrodynamic interactions between solvents and solutes. Each cube contains $N_{\rm v}$ vertex particles of mass $m_{\rm s}$ and diameter $a_{\rm v}$ arranged in a square lattice on its surface, as shown in Fig.~\ref{fig:cubemodel}(a). The vertices are connected to their nearest neighbor and the diametrically opposite one {\it via} a harmonic potential.
\begin{equation}
    U_{\rm b}(r_{ij})=\frac{k}{2}\left(r_{ij}-r_0\right)^2,
\end{equation}
where $k$ is the spring constant, $r_{ij}$ is the distance between the $i$th and $j$th particles, and $r_0$ is the equilibrium bond distance. To keep the colloidal shape nearly rigid, we set the spring constant to $k=5000\,k_{\rm B}T/a_{\rm v}^2$,\cite{ikeda:msde:2024,kobayashi:la:2022,yokoyama:sm:2023,poblete:pre:2014,wani:sm:2024} where $k_{\rm B}$ is the Boltzmann constant and $T$ is the temperature. The cube diameter is set to $d=4\,a_{\rm v}$, resulting in a lattice spacing of $4/5\,a_{\rm v}$ and $N_{\rm v}=98$ vertex particles per colloid. The size of the cube in this study is smaller than that in our previous bulk conditions.\cite{kobayashi:la:2022,ikeda:msde:2024} This condition is implemented in the present study to improve the computational efficiency while maintaining sufficient space for particle movement.

We consider suspensions of nearly rigid cubic particles with one or two hydrophobic (HO) surfaces (so-called ``patches'') while the remaining surfaces are hydrophilic (HI). Each patch consists of nine vertices (see Fig.~\ref{fig:cubemodel} (a)). The one-patch cube has a total of 6 patch arrangements, while the two-patch cube has 16; however, we only consider geometrically independent patch arrangements, as shown in Fig.~\ref{fig:cubemodel} (b--d). In the following sections, we refer to these two different realizations of two-patch cubes as types I and II, as shown in Fig.~\ref{fig:cubemodel}. We introduce the standard Lennard--Jones (LJ) potential between the vertex particles of the HO patches to mimic (effective) solvent-mediated attraction.
\begin{equation}
    U_{\rm LJ}(r_{ij}) =
    \begin{cases}
         4\varepsilon_{ij} \left[\left(\frac{a_{\rm v}}{r_{ij}}\right)^{12}-\left(\frac{a_{\rm v}}{r_{ij}}\right)^{6}\right], & r_{ij} \leq r_{\rm cut} \\
         0, & r_{ij} > r_{\rm cut}
    \end{cases}
\end{equation}
where $r_{ij}$ denotes the distance between the $i$ and $j$th particles, $\varepsilon_{ij}=1.0\,k_{\rm B}T$ denotes the interaction strength, and $r_{\rm cut}=2.5\,a_{\rm v}$ denotes the cutoff radius. The excluded volume interactions between the HI vertices and between HO and HI vertices are modeled using the standard Week--Chandler--Andersen (WCA) potential.\cite{weeks:jcp:1971}
\begin{equation}
    U_{\rm WCA}(r_{ij}) =
    \begin{cases}
        U_{\rm LJ}(r_{ij}) + \varepsilon, & r_{ij} \leq 2^{1/6}a_{\rm v}\\
         0, & r_{ij} > 2^{1/6}a_{\rm v}.
    \end{cases}
\end{equation}

The cubic particles are confined to two flat plates and experience repulsive interactions with the walls. The interactions between the nanocubes and walls are modeled as a hard impenetrable potential.\cite{lenz:sm:2009,nikoubashman:sm:2017}
\footnotesize
\begin{equation}
    \label{lj_wall}
    U_{\rm w}(x_{i}) = \begin{cases}
        \frac{2}{3}\varepsilon_{\rm w}\pi \left[\frac{2}{15}\left(\frac{a_{\rm v}}{x_{i}}\right)^9-\left(\frac{a_{\rm v}}{x_{i}}\right)^3+\frac{\sqrt{10}}{3}\right], & x_{i} \leq (\frac{2}{5})^{\frac{1}{6}}a_{\rm v} \\
        0, & x_{i} > (\frac{2}{5})^{\frac{1}{6}}a_{\rm v}
    \end{cases}
\end{equation}
\normalsize
where $\varepsilon_{\rm w}=1.0\,k_{\rm B}T$ and $x_i$ is the distance between the $i$th particle and the wall. The dimensions of the simulation box are $100\,a_{\rm v}\times 100\,a_{\rm v}\times H$, with periodic boundary conditions applied in the $x$ and $y$ directions. The walls are located at $z=\pm H/2$ with no-slip boundary conditions. We consider two slits of widths $H=15\,a_{\rm v}$ and $25\,a_{\rm v}$ with a constant volume fraction $\phi=Nd^3/L_{\rm x}L_{\rm y}H=0.0512$; thus, the total numbers of cubes are $N=200$ and 120, respectively. To observe the difference in structural properties with or without wall interactions, we also consider a bulk simulation by applying periodic boundary conditions in all directions. The edge lengths of the cubic simulation box are set to $L=80\,a_v$, with a volume fraction $\phi=0.05$ nearly equal to that of the confined systems, resulting in $N=400$ cubes. Here, we define $C=d/H$ as the ratio of the cube diameter to the slit width, which is a measure of the $C$ of the nanocube in space. Note that $C$ does not exceed $C=1$ because a slit with $H<4\,a_{\rm v}$ cannot contain cubes with $d=4\,a_{\rm v}$. To observe the effect of confinement, we include $\la M \ra$ for the bulk system ($C=0$), where $H\rightarrow\infty$. To terminate cluster growth within a finite period, we consider 1:1 binary mixtures of one- and two-patch cubes. As described below, the edge length of the cubic simulation box in the $z$ direction must be set to $L_z > H$ to guarantee no slip on the surfaces using the virtual particle technique. We then set $L_z = H + 5.0\,a_{\rm v}$.

In the MPCD technique, solvent particles are explicitly treated as point particles of mass $m_{\rm s}$ and propagated through a series of streaming and collision steps. During the streaming step, the particles are moved ballistically over a time interval $\Delta t_{\rm MPCD}$. In the subsequent collision step, the particles are packed into small cells of length $a_{\rm v}$, and their momentum is exchanged by random collisions. In this study, we use stochastic rotational dynamics, (SRD)\cite{malevanets:jcp:1999} which is a variant of the MPCD algorithm. This method involves the rotation of particles around a random axis by degree $\alpha$ during the collision step to mimic their collisional motion. The coupling of the solvent to the vertex particles is achieved through participation in the collision step (note that the MPCD solvent particles do not interact with the vertices through pair potentials). In the SRD method, the cell length $a_{\rm v}$ determines the spatial hydrodynamic resolution. In addition, the dynamic properties of the solvent are determined by its parameters: $\Delta t_{\rm MPCD}$, $\alpha$, $a_{\rm v}$, and (average) solvent number density per cell $n_{\rm c}$. Inspired by previous studies, we set the mass of the solvent particles to $m_{\rm s} = 1$, the time step to $\Delta t_{\rm MPCD} = 0.1\,\tau$ where $\tau = \sqrt{m_{\rm s}/k_{\rm B}T}a_{\rm v}$, the rotation angle to $\alpha = 130^\circ$, the cell length to $a_{\rm v} = 1$, and the number density to $n_{\rm c} = 5\,a_{\rm v}^{-3}$ to obtain a liquid-like Newtonian fluid.\cite{gompper:adv:2009,stat:prf:2019} Under these parameters, the mean free path is smaller than $a_{\rm v}$, which violates Galilean invariance. To avoid this problem, all collision cells are shifted along a randomly chosen direction before the collision steps.\cite{ihle:pre:2001}

A uniform shear flow is established by moving the two walls at a velocity $V_x$ in opposite directions with no-slip boundary conditions for the solvent particles (see Fig.~\ref{fig:cubemodel}(e)). The applied $V_x$ ranges from $10^{-3}\,a_{\rm v}/\tau$ to $10^{-1}\,a_{\rm v}/\tau$, leading to apparent shear rates $\dot\gamma$ in the range of $8.0\times 10^{-5}\,\tau^{-1}\lesssim \dot{\gamma} \lesssim 1.3\times 10^{-2}\,\tau^{-1}$ for the channel widths simulated. To ensure no slip on the surfaces under the random grid-shifting scheme, the collision cells across the walls are filled with virtual particles.\cite{lamura:epl:2001} A cell-level thermostat is employed to maintain isothermal conditions at $T = 1.0\,\varepsilon/k_{\rm B}$ and avoid viscous heating effects in the non-equilibrium simulations. All simulations are conducted using the HOOMD-blue software package (v. 2.9.6)\cite{howard:cpc:2018,anderson:jcp:2008,glaser:cpc:2015,anderson:cms:2020} with the third-party plugin ``azplugins'' (v. 0.6.2)\cite{azplugins} to introduce the wall potential given in Eq.~\eqref{lj_wall}. To achieve a better statistical analysis for each set of parameters, we conduct three independent simulations.

\section{Results and discussions}
\begin{figure}
    \centering
	 \includegraphics[width=8.5cm]{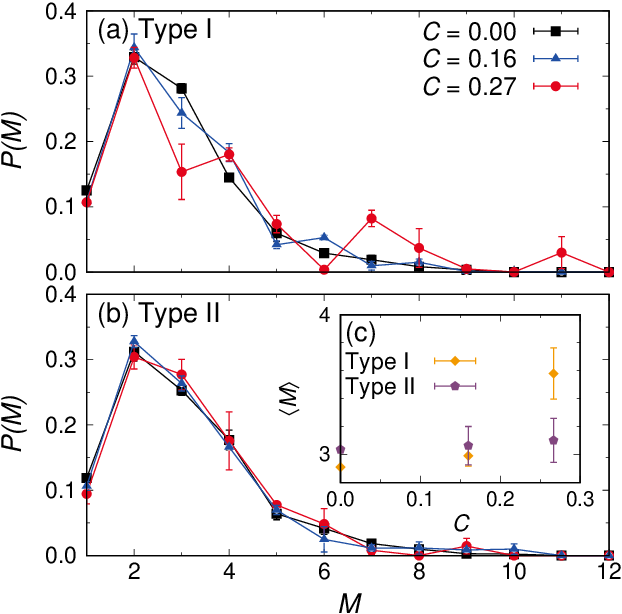}
    \caption{Cluster size distribution $P(M)$ in mixtures of one- and (a) type I and (b) type II two-patch cubes confined in nanoslits of $C=0.16$ ($H=25\,a_{\rm v}$) and $C=0.27$ ($H=15\,a_{\rm v}$) and the bulk system ($C=0.00$) at rest. (c) Mean aggregation number $\la M \ra$ as a function of the degree of confinement $C$.}
	\label{fig:ClusterDist_eq}
\end{figure}

\begin{figure}
    \centering
    \includegraphics[width=8.5cm]{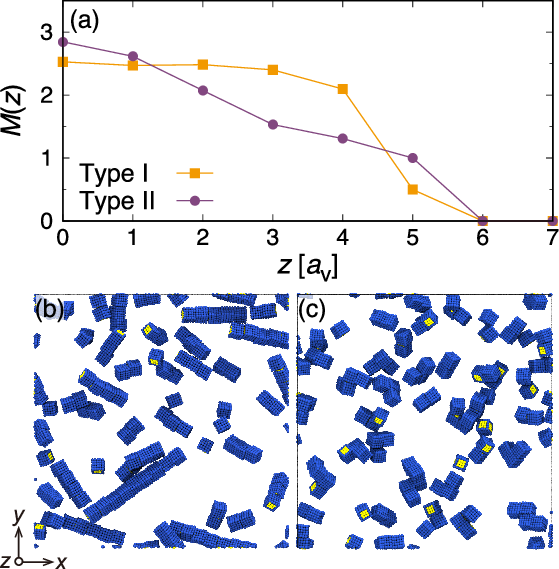}
    \caption{(a) Spatial cluster size distribution $M(z)$ of the mixtures along the direction normal to the wall surface ($z$) with $C=0.27$ ($H=15\,a_{\rm v}$). (b--c) Representative snapshots of mixtures of one- and (b) type I and (c) type II two-patch cubes confined in a slit of $C=0.27$ ($H=15\,a_{\rm v}$) from the top view in the $z$ direction.}
    \label{fig:snapshot_and_spatial_dist_eq}
\end{figure}

First, we investigated the self-assembled structures of amphiphilic patchy cubes confined to a nanoslit at rest. Cluster size distribution and mean cluster size are fundamental and crucial structural properties for amphiphilic colloidal particle suspensions because their mesoscopic structural information is strongly correlated with the resulting macroscopic transport properties, such as viscosity.
\cite{kobayashi:la:2017,kobayashi:sm:2020,kobayashi:la:2020,ikeda:msde:2024} Therefore, understanding the effect of dimensional constraint, including the interaction of the self-assembled structures with the (repulsive) walls, is important for predicting and controlling the macroscopic properties of a suspension in a nanoslit. To quantify the size and distribution of the clusters, we performed a cluster analysis using a density-based spatial clustering algorithm (DBSCAN).\cite{dbscan} Based on our previous study, nanocubes were assigned to the same cluster if the distance between the centers of mass of their HO patches was less than $2.0\,a_{\rm v}$. We then determined the aggregation number $M$ of each cluster. From these data, we calculated the mean aggregation number $\la M \ra =\sum_iP(M_i)M_i$, where $P(M)$ is the cluster size distribution, defined as the probability of finding a nanocube in an aggregate of $M$ cubes. In Fig.~\ref{fig:ClusterDist_eq}, we plotted the $P(M)$ values for mixtures of one- and two-patch cubes with different patch arrangements in a nanoslit at rest. For comparison, we also included the $P(M)$ values for the bulk system at the same $\phi$. For the bulk system, the $P(M)$ values for the different two-patch cube types are nearly identical to our measurement errors (see Fig.~\ref{fig:ClusterDist_eq}(a) and (b)), resulting in similar $\la M \ra$ for type I and type II, that was also obtained in our previous simulations.\cite{kobayashi:la:2022,yokoyama:sm:2023,ikeda:msde:2024} However, when the cubes are confined to the nanoslit, the cluster distribution exhibits a clear difference between the patch arrangements. For mixtures with type I two-patch cubes, the cluster size distribution $P(M)$ depends on $C$: the $P(M)$ for the slit of $C=0.27$ shifts to a larger $M$ compared with that for the slit of $C=0.16$, as shown in Fig.~\ref{fig:ClusterDist_eq}(a). This behavior results from the formation of perfectly rod-like aggregates in the type I two-patch cube mixture owing to the opposite arrangement of the HO patches (see Fig.~\ref{fig:cubemodel}(c)). As $H$ decreases, clusters consisting of type I two-patch cubes grow parallel to the wall surfaces (see Fig.~\ref{fig:snapshot_and_spatial_dist_eq}(b)). Therefore, the exposed HO surfaces of the clusters consistently face each other, resulting in the formation of larger ($M\geq 7$) clusters. By contrast, the $P(M)$ values for mixtures with type II two-patch cubes are nearly identical to our measurement errors, regardless of $H$, as shown in Fig.~\ref{fig:ClusterDist_eq}(b). The distinct behaviors of $P(M)$ between the different two-patch cube types are also reflected in the mean aggregation number $\la M \ra$. Figure~\ref{fig:ClusterDist_eq}(c) shows the changes in $\la M \ra$ as a function of $C$, where $C=d/H$, as defined in Section=II. Comparing $\la M \ra$ between the type I and type II mixtures at the same $C$, the dependence of this property on $C$ can be observed in the mixtures with type I two-patch cubes. By contrast, $\la M \ra$ in the mixtures of type II two-patch cubes remains approximately constant all $C$ values investigated. These differences in $P(M)$ may potentially explain the distinct (local) rheological behavior of the suspensions. Because previous studies\cite{kobayashi:sm:2020,ikeda:msde:2024} have shown that $\la M \ra$ is closely related to viscosity, the viscosity of the mixture may be controlled by controlling $H$.
\begin{figure}
    \centering
    \includegraphics[width=8.5cm]{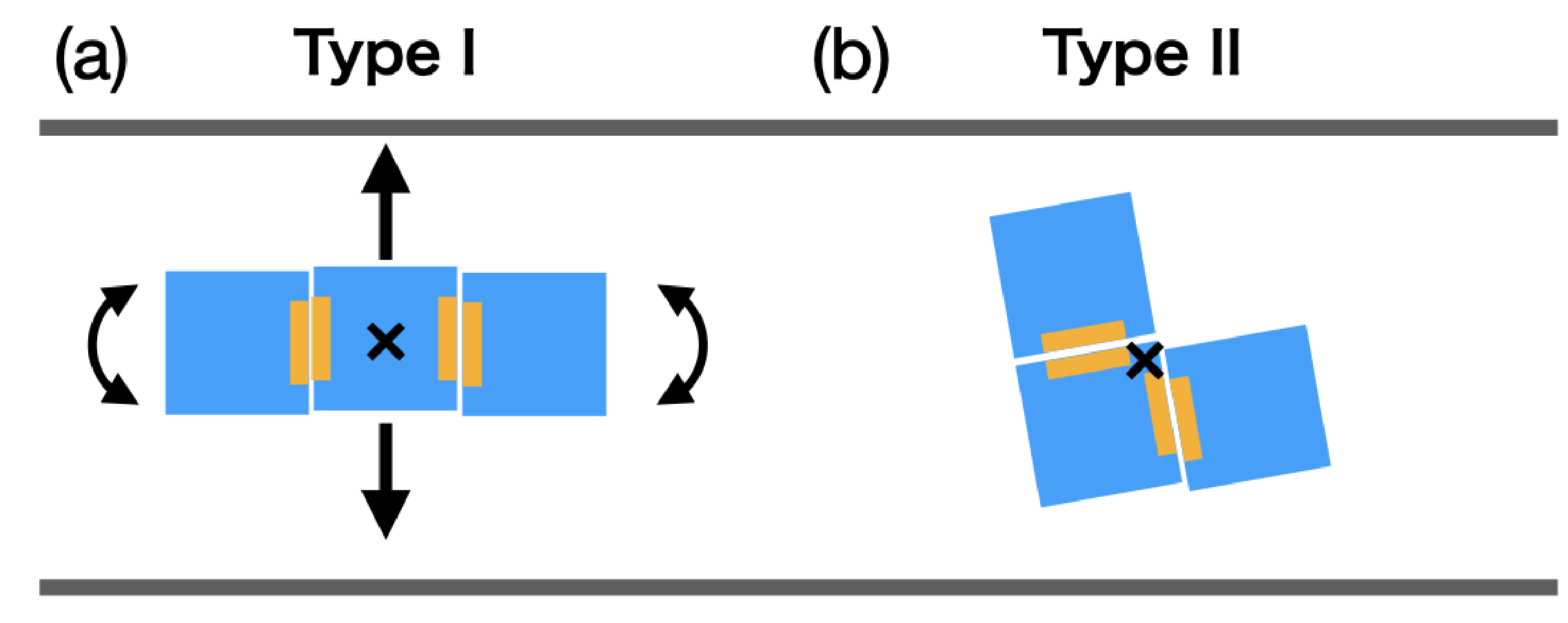}
    \caption{Schematic representation of a $M=3$ cluster of rod-like (type I) and fractal (type II) shapes in a slit.}
	\label{fig:schema_img}
\end{figure}

For a more detailed understanding, we analyzed the $M(z)$ of the mixtures along the direction normal to the wall surface ($z$) with $H=15\,a_{\rm v}$ ($C=0.27$) at rest, as shown in Fig.~\ref{fig:snapshot_and_spatial_dist_eq}(a). Here, the $M(z)$ at each $z$ position was calculated by averaging the centroids of the clusters between $z\pm 0.5\,a_{\rm v}$. Note that only the upper half of the distribution is shown because of the channel symmetry; thus, $z=0\,a_{\rm v}$ refers to the center of the slit. Based on the arrangement of the HO patches on the cubes, the type I and II two-patch cubes self-assemble into elongated rods and fractal-like structures, respectively.\cite{ikeda:msde:2024,kobayashi:la:2022,yokoyama:sm:2023} Under strong confinement ($C=0.267$), the characteristic $M(z)$ behavior appears, depending on the shape of the clusters. For mixtures with type I two-patch cubes, a uniform distribution of $M(z) \approx 2.5$ is observed over the entire range ($0\,a_{\rm v} \le z \le 4\,a_{\rm v}$) investigated. Near the wall surface ($z\ge 5\,a_{\rm v}$), the (average) spatial cluster size $M(z)$ becomes smaller than that at $0\,a_{\rm v} \le z \le 4\,a_{\rm v}$, and no clusters for $z\ge 6\,a_{\rm v}$ are observed. This finding may be attributed to the excluded volume interactions between the particles and walls. The center of mass of a single particle cannot exist in the region $z\ge(7.5-2)\,a_{\rm v}=5.5\,a_{\rm v}$. Furthermore, the centers of mass of clusters with $M>2$ cannot enter the near-wall region unless the clusters are aligned parallel to the wall surface. For mixtures with type II two-patch cubes, the peak of $M(z)$ at $z \le 1\,a_{\rm v}$ becomes more pronounced and $M(z)$ decays in the range of $z > 1\,a_{\rm v}$, which we attribute to differences in the geometric shapes of the clusters between the two patch cube types. Figure~\ref{fig:schema_img} shows a representative schematic of $M=3$ clusters of rod-like (type I) and fractal (type II) shapes in a slit with width $H=15\,a_{\rm v}$ ($C=0.27$). Rod-like clusters (Fig.~\ref{fig:snapshot_and_spatial_dist_eq}(a)) that are oriented parallel to the wall surface can move freely in the direction perpendicular to the wall surface. However, fractal-like clusters (Fig.~\ref{fig:snapshot_and_spatial_dist_eq}(b)) have three-dimensional geometries, resulting in local packing within the confined system. Consequently, the centers of mass of the fractal clusters are trapped in the central region of the slit with a noticeable peak of $M(z)$ at $z<1.0\,a_{\rm v}$.
\begin{figure}
    \centering
    \includegraphics[width=8.5cm]{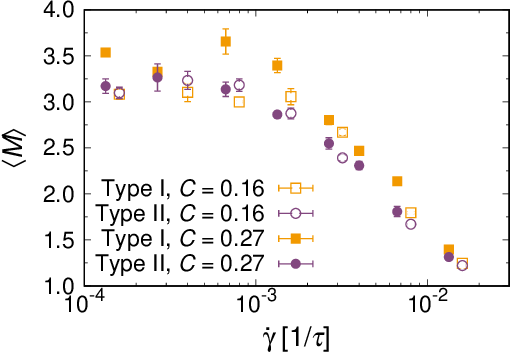}
    \caption{Mean aggregation number $\la M \ra$ as a function of the shear rate $\dot\gamma$ for mixtures of one- and two-patch cubes confined in slits.}
	\label{fig:MAN_vs_shear}
\end{figure}
\begin{figure}
    \centering
    \includegraphics[width=8.5cm]{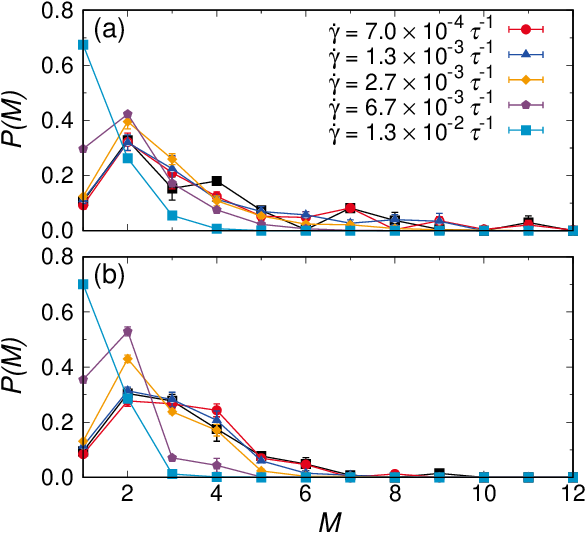}
    \caption{Cluster size distribution $P(M)$ in sheared mixtures of one- and (a) type I and (b) type II two-patch cubes confined in a slit of $C=0.27$ ($H=15\,a_{\rm v}$).}
	\label{fig:ClusterDist}
\end{figure}
\begin{figure}
    \centering
    \includegraphics[width=8.5cm]{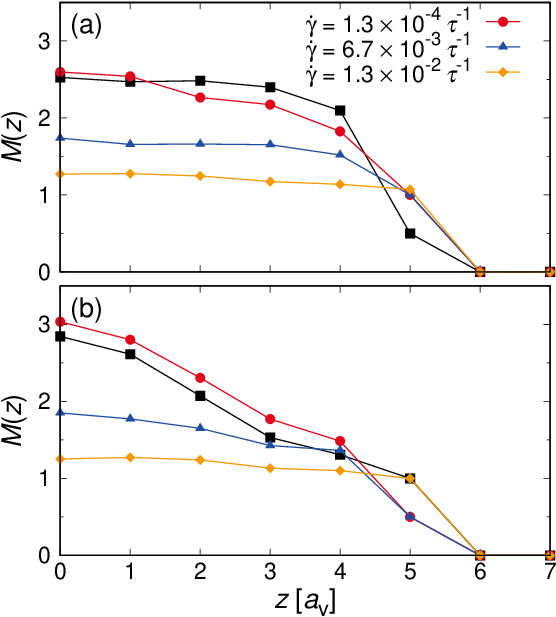}
    \caption{Spatial cluster size distribution $M(z)$ in mixtures of (a) type I and (b) type II two-patch cubes along the direction normal to the wall surface ($z$) with $C=0.27$ ($H=15\,a_{\rm v}$) under various shear rates $\dot\gamma$. The black lines represent the data at the equilibrium state.}
	\label{fig:spatial_dist}
\end{figure}

Next, we discuss the steady-state structural morphologies confined to the nanoslit under shear. Figure~\ref{fig:MAN_vs_shear} shows the changes in $\la M \ra$ as a function of $\dot\gamma$. Comparing $\la M \ra$ between the type I and type II mixtures at the same $\dot\gamma$, we find that type I mixtures have larger $\la M \ra$ than type II mixtures for two different $C$ over the entire range of $\dot\gamma$ investigated. This behavior is peculiar to the confined systems. For the bulk system, the bonding energy per cube $E_{\rm LJ}/M$ for aggregates composed of $M$ type II cubes is consistently higher than that for aggregates of type I cubes. Thus, clusters composed of type II cubes are more resistant to shear than clusters of type I cubes.\cite{ikeda:msde:2024} We anticipate that the parallel orientation of the rodlike clusters consisting of type I two-patch cubes to the wall surfaces can be enhanced by restricting the three-dimensional free rotational motion of the clusters in a confined slit space. Such a parallel alignment along the flow direction would reduce the shear-induced collision of each cluster.

Figure~\ref{fig:ClusterDist} shows the $P(M)$ for sheared mixtures of one- and two-patch cubes confined in a slit of width $H=15\,a_{\rm v}$. In mixtures with type I two-patch cubes under weak shear ($7.0\times 10^{-4}\,\tau^{-1} \le \dot\gamma \le 2.7\times 10^{-3}\,\tau^{-1}$), the peak of $P(M)$ at $6\le M \le 8$ remains roughly unchanged (see Fig.~\ref{fig:ClusterDist}(a)), indicating the existence of long elongated rods even in sheared systems. However, in the case of type II two-patch cubes, the peak of $P(M)$ at $M\ge 6$ is much less pronounced (Fig.~\ref{fig:ClusterDist}(b)). Furthermore, $P(M)$ at $M=3$ and $M=4$ remains nearly constant for $\dot\gamma \le 2.7\times 10^{-3}\,\tau^{-1}$ and exceeds that of the type I mixtures, reflecting the high stability of clusters with fractal shapes against shear. As $\dot\gamma$ increases, the peak of $P(M)$ shifts to lower $M$ values for both mixtures, indicating a gradual breakup of the clusters with $\la M \ra \rightarrow 1$ at $\dot\gamma \gtrsim 10^{-2}\,\tau^{-1}$. 

Figure~\ref{fig:spatial_dist} shows the $M(z)$ of mixtures confined in a slit of width $H=15\,a_{\rm v}$ under various $\dot\gamma$. For the mixtures with type I two-patch cubes (Fig.~\ref{fig:spatial_dist}(a)), $M(z)$ slightly increases around the center of the slit ($0\,a_{\rm v}\le z \le 1\,a_{\rm v}$) but decreases in the region of $z \ge 1\,a_{\rm v}$ under weak shear ($\dot\gamma = 1.3\times 10^{-4}\,\tau^{-1}$) compared with the equilibrium structures. For the mixtures with type II two-patch cubes (Fig.~\ref{fig:spatial_dist}(b)), we observe (weak) shear-induced cluster growth; $M(z)$ completely increases and is highest at the center of the slit. For both mixtures, as $\dot\gamma$ further increases, $M(z)$ shows a more uniform distribution. We also find that $M(z)$ increases to $M\approx 1.25$ at $z=5\,a_{\rm v}$ because of the increase in free particles in the system owing the gradual breakup of the clusters.

\section{Summary and Conclusions}
MD simulations coupled with MPCD were performed to investigate the structure formation and dynamics of amphiphilic cubes in nanoslits. We examined two types of dilute mixtures of one- and two-patch cubes in a 1:1 ratio with different HO patch arrangements at a constant volume fraction. To study the effect of confinement, we considered two different $H$ as well as bulk conditions. The confined nanocubes experienced a shear flow induced by the walls moving at a constant $V_x$.

The resulting shapes of the clusters depended on the arrangement of the HO patches; rod- and fractal-like aggregates were obtained in two-patch cubes with opposite and adjacent arrangements, respectively. We also observed that the cluster sizes of the rod-like aggregates depended on $C$, whereas those of the fractal-like aggregates remained nearly constant regardless of $C$. The $\la M \ra$ of the rod-like aggregates increased with decreasing $H$, accompanied by a shift to a larger $M$ in the cluster distribution. This increase was related to the spatial arrangement of the clusters under confinement. The rod-like clusters grew parallel to the wall with the HO planes consistently facing each other, leading to the formation of more elongated rods with $M>6$, which was not observed in the bulk condition. By contrast, the distributions and cluster sizes of the fractal-like aggregates remained nearly identical over the entire range of $H$ investigated.

To further investigate the spatial arrangement of the clusters under strong confinement, we analyzed $M(z)$ along the direction normal to the wall surface. At rest, the rod-like aggregates had a uniform size distribution; however, the cluster size of the fractal-like aggregates was largest at the center of the slit and became smaller than that of the rod-like aggregates farther away from the center. This finding was attributed to differences in the geometric shapes of the clusters between the two-patch cube types: fractal-like clusters of two- or three-dimensional shapes were trapped within the confined geometry, whereas rod-like clusters (one-dimensional shapes) were oriented parallel to the walls and moved freely in the direction normal to the walls. 

We also studied the structural changes in the clusters under shear. In the weakly sheared mixture, the $\la M \ra$ of the rod-like aggregates was larger than that of the fractal-like aggregates. Such behavior is unique to self-assembled structures with a confined geometry because fractal-like aggregates are more resistant to flow than rod-like aggregates owing to their higher bonding energy in the bulk state. These differences in $P(M)$ may potentially explain the distinct (local) rheological behavior of the suspensions. Because previous studies\cite{kobayashi:sm:2020,ikeda:msde:2024} have shown that $\la M \ra$ is closely related to viscosity, the viscosity of the mixture may be controlled by controlling $H$.

Our study focused only on the hydrophilic (HI) (repulsive) interactions of the walls in dilute suspensions. Modification of the wall surfaces with HO materials may lead to different self-assembled morphologies compared with those observed in the presence of purely repulsive interactions.\cite{kobayashi:sm:2015,nikoubashman:sm:2017,kobayashi:msde:2019} Moreover, anisotropic wall interactions, such as those found in Janus-like nanotubes,\cite{sun:pnas:2016,kobayashi:ms:2017,tsujinoue:la:2020} slits,\cite{baran:ao:2023} and the stripe pattern of HI and HO wall surfaces,\cite{arai:la:2012} may yield more complex self-assembled structures. In addition to wall interactions, future work in this area should consider the density of nanocubes in the suspension. In principle, a higher volume fraction of amphiphilic nanocubes could form larger clusters with highly orientationally ordered structures such as nematic or smectic structures, similar to lyotropic liquid crystals. We plan to study such systems in the future.


\section*{Author contributions}
\textbf{Takahiro Ikeda}: Conceptualization, formal analysis, investigation, methodology, visualization, writing -- original draft. \textbf{Yusei Kobayashi}: Conceptualization, project administration, supervision, writing -- review \& editing. \textbf{Masashi Yamakawa}: Supervision, writing – review \& editing.

\section*{Conflicts of interest}
The authors have no conflicts to disclose.

\section*{Data Availability}
The data that support the findings of this study are available from the authors upon reasonable request.

\begin{acknowledgements}
This study was supported by a grant from the Japan Science and Technology Agency (JST) and Support for Pioneering Research Initiated by the Next Generation (SPRING) Grant Number JPMJSP2107. Y.K. acknowledges JSPS KAKENHI Grant No. JP24K17216 and the support of KIT Grants-in-Aid for Early-Career Scientists.
\end{acknowledgements}

\bibliography{cite} 

\end{document}